\documentclass{aa}
\usepackage{psfig,txfonts}

\begin{document}

\title{Yet another galaxy identification for an\\ 
ultraluminous X-ray source\thanks{Based on observations collected
at the Bologna Astronomical Observatory in Loiano, Italy, and
at the Astronomical Observatory of Asiago, Italy.}}

\author{
N. Masetti\inst{1},
L. Foschini\inst{1,2},
L. C. Ho\inst{3},
M. Dadina\inst{1},
G. Di Cocco\inst{1},
G. Malaguti\inst{1} and
E. Palazzi\inst{1}
}

\institute{
Istituto di Astrofisica Spaziale e Fisica Cosmica --- Sezione di Bologna, 
CNR, via Gobetti 101, I-40129, Bologna (Italy)
\and
INTEGRAL Science Data Centre, Chemin d'Ecogia 16, CH-1290 Versoix 
(Switzerland)
\and
The Observatories of the Carnegie Institution of Washington, 813 Santa 
Barbara Street, Pasadena, CA 91101 (USA)
}

\offprints{N. Masetti (\texttt{masetti@bo.iasf.cnr.it)}}
\date{Received 22 April 2003; Accepted 22 May 2003}

\abstract{We report on the identification of the optical and infrared 
counterparts of the ultraluminous X-ray source (ULX)
XMMU~J121214.5+131248
(NGC4168-ULX1). The optical spectrum yields a redshift of $z=0.217$,
which implies that the ULX is not associated with the nearby galaxy
NGC~4168, but rather with a background object.
Optical spectral line ratios and the spectral energy distribution 
constructed from the available data indicate that the source
is likely a starburst nucleus.
\keywords{Galaxies: distances and redshifts ---
Galaxies: photometry --- Galaxies: starburst}}

\titlerunning{ULX identification in NGC~4168}
\authorrunning{N. Masetti et al.}

\maketitle

\section{Introduction}

\vspace{-.2cm}
Ultraluminous X-ray sources (ULXs), i.e. off-nuclear sources with
luminosities well above the Eddington limit for a typical neutron star,
are now known to populate about 30\% of nearby galaxies (see
van~der~Marel 2003 for a review). Despite much effort, little is
presently known about the nature of these sources, particularly since
the counterparts at other wavelengths are difficult to find. Indeed,
these sources are often superposed against regions of high surface
brightness in their host galaxy. Thus, most of the optical
identifications are obtained by spatial coincidence of the counterparts
and comparison with broadband spectral characteristics of known stars
in the host (Liu et al. 2002; Wu et al. 2002; Zezas et al. 2002;
Zampieri et al. 2003). In other cases, it has been possible to study only 
the nearby environment in the ULX host (Pakull \& Mirioni 2002; Wang 2002; 
Roberts et al. 2003).
In one case (NGC4698-ULX1) the optical spectral features allowed a clear
identification as a background BL Lac object (Foschini et al. 2002a).

Here we report the identification of the nature of another ULX. The
counterparts in the infrared and optical bands were found, and
spectroscopy revealed clear Balmer and forbidden transition emission
lines. The derived redshift of $z=0.217$ indicates a background galaxy
in this case also.

\vspace{-.3cm}
\section{The source: NGC4168-ULX1}

\vspace{-.2cm}
NGC~4168 is an E2 elliptical galaxy located in the Virgo cluster
($d=16.8$~Mpc). It hosts an active galactic nucleus (AGN), classified
as a Seyfert 1.9 by Ho et al. (1997). The galaxy was observed on 4
December 2001 using the European Photon Imaging Camera (EPIC) on board
the \emph{XMM-Newton} satellite (see Foschini et al. 2002b; for the
X-ray part, we refer in the following to the results obtained in this
paper, unless explicitly stated). EPIC is composed of two instruments:
the PN-CCD camera (Str\"uder et al. 2001) and two MOS-CCD detectors
(Turner et al. 2001). The effective exposure time was $17.4$~ks.

One ULX was found apparently associated with NGC~4168, being inside the 
galaxy's $D_{25}$ ellipse, at $45''$ from the optical centre of the galaxy. 
The ULX has coordinates (J2000) $\alpha = 12^{\rm h} 12^{\rm m}
14\fs5$ and $\delta = +13^{\circ} 12\arcmin 48\arcsec$, with an
uncertainty radius of $4''$.

The counts were not sufficient to extract a spectrum, so we could only
convert to physical units using the count rates derived from the 
\emph{eboxdetect} task of the XMM-SAS software (v. 5.2).
We found a count rate of $4.0 \pm 0.7$~counts s$^{-1}$ in the 0.5--10
keV band, which corresponds to a flux of $(1.8\pm 0.3)\times
10^{-14}$~erg~cm$^{-2}$~s$^{-1}$ adopting a conversion factor of
$3\times 10^{11}$~counts~cm$^2$ erg$^{-1}$, which in turn was derived
by using a power-law model with $\Gamma=2.0$ and an average Galactic
column density of $N_{\rm H} = 3\times 10^{20}$~cm$^{-2}$. This value
for $\Gamma$ is common among the ULXs found with \emph{XMM-Newton}.
This flux value was corrected according to the energy encircled fraction
(Ghizzardi 2001). Correction for vignetting (Lumb 2002) was not
applied because the source is close to the center of the field of view
($<$$2\arcmin$). 
At the distance of NGC~4168 the resulting $0.5-10$~keV luminosity is
$6\times 10^{38}$~erg s$^{-1}$ (for a discussion on the luminosity
threshold of ULXs, see Foschini et al. 2002c).

\vspace{-.3cm}
\section{Archival optical and near-infrared data}

\vspace{-.2cm}
A single optical object with coordinates (J2000) $\alpha = 12^{\rm h}
12^{\rm m} 14\fs60$ and $\delta = +13^{\circ} 12\arcmin 47\farcs$8 (i.e., 
1$\farcs$4 from the \emph{XMM-Newton} position, thus well within the X-ray 
error box) is clearly visible on the Digitized Sky Survey
(DSS)\footnote{\texttt{http://archive.eso.org/dss/dss/}} observation of
NGC~4168, originally made with the 48-inch Schmidt telescope at Palomar
Observatory on 14 April 1955. 
We thus consider it as the optical counterpart of NGC4168-ULX1.

The source is also present in the US Naval Observatory (USNO) B1.0 Catalog
(Monet et al. 2003) with the identification number $1032-0222128$. The
magnitudes in the different bands are $B1=19.1$, $B2=17.8$, $R1=18.1$,
$R2=18.2$, and $I=17.8$. The $B1$ and $R1$ magnitudes refer to the Palomar
Observatory Sky Survey I (POSSI), performed between 1949 and 1965. The
$B2$, $R2$ and $I$ magnitudes are measured from the Palomar Observatory
Sky Survey II (POSSII), performed from 1985 to 2000.

\begin{figure}[t!]
\begin{center}
\psfig{file=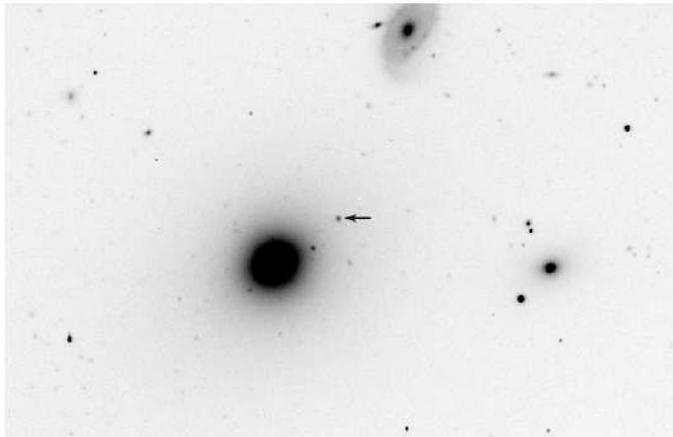,width=9cm}
\end{center}
\vspace{-1cm}
\caption{$R$-band image (exposure time: 10 minutes) of NGC4168-ULX1 
acquired in Loiano starting at 23:52 UT of 4 March 2003. North is up
and East to the left. The field size is 8$'$$\times$5$'$. The arrow 
indicates the position of the source}
\end{figure}

Two Micron All Sky Survey 
(2MASS)\footnote{\texttt{http://www.ipac.caltech.edu/2mass/}}
near-infrared (NIR) observations were performed on 5 April 2000, and the
source was detected in all the three bands ($JHK$). From the Second
Point Source Catalog, we get $J$ = 16.8$\pm$0.2, $H$ = 16.1$\pm$0.2, 
and $K$ = 15.3$\pm$0.1.

\vspace{-.3cm}
\section{Optical observations at Loiano and Asiago}

\vspace{-.2cm}
\subsection{Spectra}

\vspace{-.2cm}
Six medium-resolution optical spectra of the ULX were acquired between
21:23 UT of 4 March and 03:24 UT of 5 March 2003 in Loiano (Italy)  
with the Bologna Astronomical Observatory $1.52$~metre ``G.D. Cassini''
telescope plus BFOSC, for a total exposure time of 3 hours. One single
1-hour medium-resolution spectrum was also obtained at the ``L.  
Rosino'' Astronomical Observatory of Asiago (Italy), starting at 01:02
UT of 26 February 2003, with the $1.82$~metre ``Copernicus'' telescope
plus AFOSC. The Cassini telescope was equipped with a $1300\times1340$
pixels EEV CCD. The Copernicus telescope mounted instead a
$1100\times1100$ pixels SITe CCD. In both cases, Grism \#4 and a slit
width of $2''$ were used, providing nominal spectral coverages
3500-8500 \AA~(Loiano) and 4000-8000 \AA~(Asiago). The use of this
setup secured a final dispersion of $4.0$~\AA/pix and $4.3$~\AA/pix for
the spectra acquired in Loiano and Asiago, respectively.

Spectra, after correction for flat-field, bias and cosmic-ray
rejection, were background subtracted and optimally extracted (Horne
1986) using IRAF\footnote{\texttt{http://iraf.noao.edu/}}.
Wavelength calibration of
the Loiano spectra was performed using He-Ar lamps, while the one
acquired in Asiago was calibrated with Cd-Hg-Ne lamps. Spectra taken
in Loiano were then flux-calibrated by using the spectrophotometric
standard Hiltner 600 (Hamuy et al. 1992, 1994) and finally stacked
together. Correction for slit losses was also applied to the continuum.
For the Asiago spectrum no spectroscopic standard was available. 
Wavelength calibration was checked by using the positions of background
night sky lines; the error was $\sim$0.5~\AA~for both spectra.

\vspace{-.3cm}
\subsection{Imaging}

\vspace{-.2cm}
Between 23:45 UT of 4 March and 01:28 UT of 5 March 2003 $UBVRI$
photometry was also acquired in Loiano with BFOSC under an average 
seeing of 1$\farcs$5.
The EEV CCD, with a scale of 0$\farcs$58/pix, secured a field of
12$\farcm$6$\times$12$\farcm$6. Images were corrected for bias and
flat-field in the usual fashion and calibrated using the Rubin 149
field (Landolt 1992); the calibration accuracy is better than 3\% in $BVR$
and better than 5\% in $UI$. The source (in Fig. 1) is well detected in 
all bands.

Galactic absorption in the optical and NIR bands along the direction of
the source was evaluated using the Galactic dust infrared maps by
Schlegel et al. (1998); from these data we obtained a color excess
$E(B-V)$ = 0.036 mag. By applying the relation of Cardelli et al. (1989),
we derived the following magnitudes for Galactic extinction correction: 
$A_U$ = 0.17, $A_B$ = 0.15, $A_V$ = 0.11, $A_R$ = 0.08,
$A_I$ = 0.07, $A_J$ = 0.03, $A_H$ = 0.02, and $A_K$ = 0.01.

\begin{figure}[t!]
%\begin{center}
\hspace{-.5cm}
\psfig{file=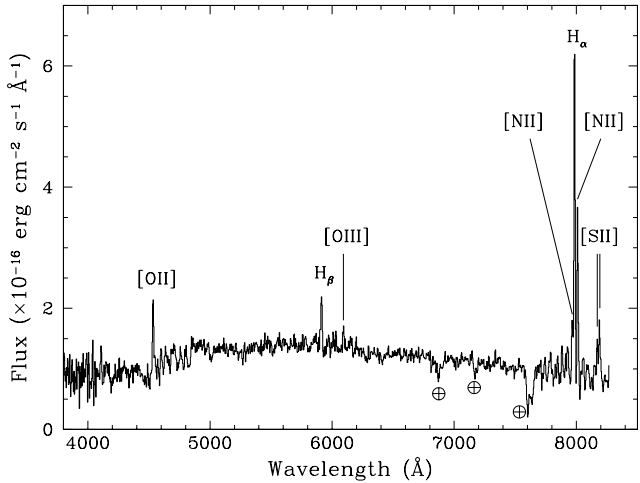,width=10cm,angle=270}
%\end{center}
\vspace{-1cm}
\caption{Average optical spectrum of NGC4168-ULX1 taken with the Cassini
telescope at Loiano. The main spectral features are labeled. These
allowed us to determine the redshift of the source as $z$ = 0.217. The
symbol $\oplus$ indicates atmospheric telluric features}
\end{figure}

\begin{table}[t!]
\caption[]{Fluxes (in units of 10$^{-15}$ erg s$^{-1}$ cm$^{-2}$) of 
the emission lines detected in the spectrum acquired in Loiano. Values 
are corrected for Galactic absorption}
\begin{center}
\vspace{-.3cm}
\begin{tabular}{lr}
\noalign{\smallskip}
\hline
\noalign{\smallskip}
\multicolumn{1}{c}{Line} & \multicolumn{1}{c}{Flux} \\
\noalign{\smallskip}
\hline
\noalign{\smallskip}

$[$O {\sc ii}$]$ $\lambda$3727  & 1.51$\pm$0.15 \\
H$_\beta$                        & 1.1$\pm$0.2 \\
$[$O {\sc iii}$]$ $\lambda$5007 & 0.35$\pm$0.09 \\
$[$N {\sc ii}$]$ $\lambda$6548  & 0.90$\pm$0.18 \\
H$_\alpha$                       & 6.4$\pm$0.3 \\
$[$N {\sc ii}$]$ $\lambda$6583  & 2.90$\pm$0.15 \\
$[$S {\sc ii}$]$ $\lambda$6716 & 0.63$\pm$0.13 \\
$[$S {\sc ii}$]$ $\lambda$6731 & 0.86$\pm$0.17 \\

\noalign{\smallskip}
\hline
\noalign{\smallskip}
\end{tabular}
\end{center}
\end{table}

\vspace{-.3cm}
\section{Results}

\vspace{-.2cm}
The average spectrum taken at Loiano (Fig. 2) shows a number of 
emission features 
that can be readily identified with typical optical nebular lines. These 
include [O~{\sc ii}] $\lambda$3727, H$_\beta$, [O~{\sc iii}] $\lambda$5007, 
H$_\alpha$, [N~{\sc ii}] $\lambda\lambda$6548, 6583, and 
[S~{\sc ii}] $\lambda\lambda$6716, 6731.
All identified emission lines yield a redshift of $z=0.217$.
This result is, in hindsight, not surprising, as the images acquired in 
Loiano show that the field is well populated by galaxies of various 
sizes and brightnesses, with very few foreground stars.
The Asiago spectrum, albeit of lower S/N due to its shorter exposure, shows
the presence of H$_\beta$ and [O~{\sc ii}] $\lambda$3727 at the same
redshift, thus confirming the result of the Loiano observations.
Unfortunately, the actual useful spectral range secured by this spectrum
has its red end at 7800 \AA; therefore it does not cover the region
containing H$_\alpha$. Table 1 reports the emission-line fluxes as determined 
from the Loiano spectrum, dereddened for Galactic absorption. Given the 
limited S/N and resolution of the spectrum, no correction for starlight 
contamination (e.g., Ho et al. 1993, 1997) was attempted, but this 
does not strongly affect any of our conclusions.

Magnitudes of the source in the Loiano images were measured through
aperture photometry, since its profile is significantly larger than the
image PSF (2$\farcs$6 versus 1$\farcs$5) and, as measured on the $U$-band
frame, is possibly elongated in the NW-SE direction. Using an aperture radius 
of 4.5 pixels (corresponding to 2$\farcs$6), we found the following optical
magnitudes (not corrected for Galactic absorption):
$U$ = 19.49$\pm$0.08, $B$ = 19.78$\pm$0.04, $V$ = 18.69$\pm$0.03,
$R$ = 18.33$\pm$0.03, and $I$~=~17.68$\pm$0.05.
The broad-band spectral energy distribution (SED) of the source,
constructed with data from \emph{XMM-Newton} (X-ray), Loiano (optical), and 
2MASS (NIR), is shown in Fig. 3.
The optical-NIR data points were corrected for Galactic absorption and
converted into flux densities using the tables by Fukugita et al. (1995)
for the optical and by Bersanelli et al. (1991) for the NIR.  A 5\% systematic 
error was added in quadrature to account for the uncertainties in the 
magnitude-to-flux conversion factors (Fukugita et al. 1995).
No correction for any possible further absorption along the line of
sight produced by the halo of NGC~4168 was included; this, however,
should not be high as the source is well detected in the $U$ band.

\begin{figure}[t!]
%\begin{center}
\hspace{-.5cm}
\psfig{file=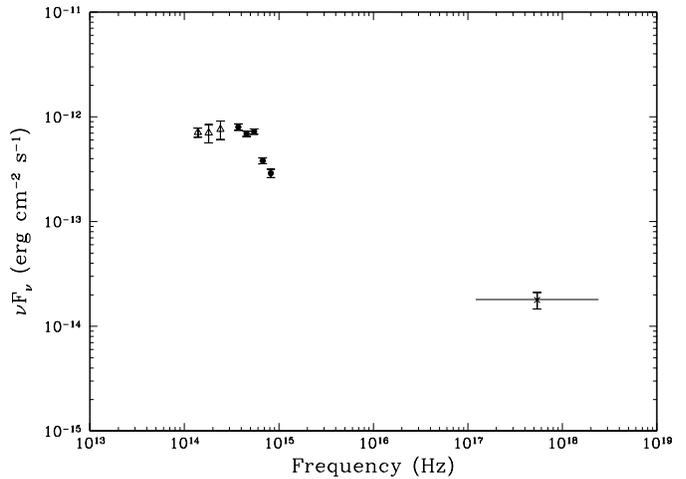,width=9.5cm,angle=270}
%\end{center}
\vspace{-1cm}
\caption{Broad-band SED of NGC4168-ULX1 composed of data from 
\emph{XMM-Newton} (X-ray, cross), Loiano (optical, filled dots), and 2MASS 
(NIR, open triangles).  Optical and NIR data were corrected for the Galactic 
absorption along the source line of sight. The presence of a peak in the SED 
across the optical-NIR domain is apparent}
\end{figure}

\vspace{-.3cm}
\section{Discussion}

\vspace{-.2cm}
So, yet another putative ULX is found to be a background source. Assuming
a cosmology with $H_{\rm 0}$ = 65 km s$^{-1}$ Mpc$^{-1}$,
$\Omega_{\Lambda}$ = 0.7 and $\Omega_{\rm m}$ = 0.3, we find that the
luminosity distance to this source is $d_L$ = 1.16 Gpc, and that its
X-ray luminosity is $2.9\times 10^{42}$~erg s$^{-1}$ in the 0.6--12 keV
rest-frame energy range. The angular size of the source translates into a 
linear diameter of about 30 kpc at $z$ = 0.217.

The measured value for the luminosity is $\sim$100--1000 times less 
that of ``classical'' active galactic nuclei (AGNs). Indeed, visual 
inspection of the optical spectrum in Fig. 2 suggests that the lines are 
due to stellar photoionization, rather than to an AGN. The diagnostic 
line ratios [N~{\sc ii}]/H$_\alpha$, [S~{\sc ii}]/H$_\alpha$, and 
[O~{\sc iii}]/H$_\beta$, together with the nondetection of substantial 
[O~{\sc i}] $\lambda$6300 emission, place this source in the regime of 
metal-rich giant extragalactic H~{\sc ii} regions or starburst nuclei 
(Ho et al. 1993, 1997).

Inspection of the SED of the source (Fig. 3) shows that a peak in the
optical-NIR domain is present, followed by a drop across the $U$ and $B$ 
bands.  Given the dominance of the stellar continuum in the spectrum (Fig. 2), 
most of the optical and NIR light comes from the integrated stellar
emission from the galaxy.  Although the S/N of the spectrum is insufficient 
to place strong constraints on the age and metallicity of the stellar 
population, they are not inconsistent with that of an evolved population.
The optical-NIR SED also supports this conjecture.  The X-ray emission, 
by contrast, is likely to be associated with the star-forming 
regions that give rise to the optical line emission.  This may consist of 
a single starburst nucleus or multiple off-nuclear sources. An X-ray
luminosity of $\approx$10$^{42}$~erg s$^{-1}$ is high, but not unusual,
for a starburst (David et al. 1992). 

The strength of the optical emission lines, after accounting for 
internal reddening, can be used to estimate the star formation rate (SFR) and 
metallicity. Assuming an intrinsic Balmer decrement of H$_\alpha$/H$_\beta$ = 
2.86 (Osterbrock 1989) and the extinction law of  Cardelli et al. (1989), 
the observed H$_\alpha$/H$_\beta$ = 5.82 implies an internal reddening of 
$E(B-V)$ = 0.72 mag. Following Kennicutt (1998), we determine a SFR of 
43$\pm$5 $M_\odot$ yr$^{-1}$ from the reddening-corrected H$_\alpha$ 
luminosity of $5.5 \times 10^{42}$ erg s$^{-1}$. Similarly, the 
[O {\sc ii}] luminosity yields a SFR of 80$\pm$20 $M_\odot$ yr$^{-1}$. 
Thus, to within a factor of $\sim$2 and within the uncertainties, the two 
estimates are consistent.

The optically derived SFR, on the other hand, is significantly lower
than that predicted from the total X-ray luminosity.  From the
calibration of Ranalli et al. (2003), a 0.5--10 keV luminosity of
3$\times$10$^{42}$~erg s$^{-1}$, if entirely attributed to star
formation, requires a SFR of $\approx$300 $M_\odot$ yr$^{-1}$, a factor
of $\sim$5 higher than our estimates based on optical line emission.  
There are two possible explanations for this apparent discrepancy.  
First, the optically derived SFR may be an underestimate of the total
SFR if star formation is more extended than the 2$''$-wide slit: 
indeed, in case the distribution of star-forming regions is wider than 
that of the optical continuum emission, the correction for slit losses 
underestimates the actual total flux of emission lines. 
By contrast, the X-ray emission integrates over a larger area. Second,
only a portion ($\sim$20\%) of the X-ray emission originates from star
formation, with the dominant fraction coming from an AGN. Although the
optical spectrum shows no signs of nonstellar activity, we cannot
exclude the presence of a low-level AGN, especially one that is heavily
obscured. Both of these possibilities can be tested with further
observations.

The detection of [O {\sc ii}], [O {\sc iii}], and H$_\beta$ also allows us
to infer the gas-phase oxygen abundance. Following Kobulnicky et al.
(1999), the $R_{\rm 23}$ parameter gives 12 + log (O/H) = 7.6 or 8.9.
Considering the intrinsic luminosity of the source (rest-frame $M_B
\approx$ $-$21.5 mag) and its [O {\sc iii}]/[N~{\sc ii}] ratio, the larger
of the two values is likely to be the correct one (see Kobulnicky et al.
1999), pointing to a basically solar oxygen abundance.

The comparison of the USNO archival magnitudes with the results of
the optical photometry acquired in Loiano seems to suggest long-term
variability for the source. However, we caution the reader that it is not
infrequent to find discrepancies as large as 0.6 mag (see e.g. Masetti et
al. 2003) between the automatically extracted archival magnitudes and the
actual ones, calibrated through Landolt (1992) fields. This is most likely
due to the fact that magnitude extraction from archival images was
performed by using automatic pipelines and was (for the oldest surveys)
applied to photographic plates. 

In the minisurvey in Foschini et al. (2002b) 18 ULXs in 10 nearby galaxies
were detected. According to the statistics from the Lockman hole studies
(Hasinger et al. 2001) we expect to find 3.2 background sources with 2--10
keV flux higher than $10^{-14}$~erg~cm$^{-2}$~s$^{-1}$ in an area
corresponding to the sum of all the $D_{25}$ ellipses of the considered
galaxies. If we also take into account the case reported in Foschini et
al. (2002a), with the present identification we already found 2 background
objects among the 18 ULX candidates: so, we are close to the statistical
limit by assuming that the Lockman Hole studies are valid over a wide
range of the sky. Albeit the two selected cases might suffer from an
observational bias (they are quite bright and outside the main body of the
parent galaxy), additional detection of other background objects in the
minisurvey should be considered with care.

In conclusion, we wish to remark that this case, as that of the ULX in
NGC~4698 reported by Foschini et al. (2002a), has been instructive:
despite the availability of broad-band multiwavelength data, they have
proven insufficient to disclose the true nature of the source.  The most
crucial piece of information is optical spectroscopy of sufficient quality
to reveal clear spectral features, which simultaneously provide redshift
information and spectral diagnostics.

\vspace{-.3cm}
\begin{acknowledgements}
We wish to thank L. Zampieri for the coordination of the Asiago Service
Mode observations, and P. Grandi and M. Cappi for useful discussions. We
also thank R. Gualandi for the night assistance in
Loiano and H. Navasardyan for having performed the observations in Asiago
as a Service Mode run. This research has made use of the NASA Astrophysics
Data System Abstract Service, of the NASA/IPAC Extragalactic Database
(NED), and of the NASA/IPAC Infrared Science Archive, which are operated
by the Jet Propulsion Laboratory, California Institute of Technology,
under contract with the National Aeronautics and Space Administration.
\end{acknowledgements}

\end{document}